\documentclass[conference]{IEEEtran}
\usepackage[a4paper, margin=1in]{geometry}
\IEEEoverridecommandlockouts
\usepackage{cite}
\usepackage{amsmath,amssymb,amsfonts}
\usepackage{algorithmic}
\usepackage{graphicx}
\usepackage{textcomp}
\usepackage{xcolor}
\usepackage{subcaption}
\usepackage{multirow}
\usepackage{booktabs}
\usepackage{float}
\begin{document}
\title{FaultGene: LLM-Assisted Fault-Augmented Test Generation for Hardware-in-the-Loop Validation of Autonomous Driving Systems\\}

\author{\IEEEauthorblockN{Chaoran Yuan}
\IEEEauthorblockA{\textit{Center for Embedded and Cyber-Physical Systems} \\
\textit{University of California, Irvine}\\
Irvine, USA \\
chaory2@uci.edu}\\

\IEEEauthorblockN{Fadi kurdahi}
\IEEEauthorblockA{\textit{Center for Embedded and Cyber-Physical Systems} \\
\textit{University of California, Irvine}\\
Irvine, USA \\
kurdahi@uci.edu}

}

\author{\IEEEauthorblockN{1\textsuperscript{st} Chaoran Yuan}
\IEEEauthorblockA{\textit{Center for Embedded and Cyber-Physical Systems} \\
\textit{University of California, Irvine}\\
Irvine, USA \\
chaory2@uci.edu}
\and
\IEEEauthorblockN{2\textsuperscript{nd} Fadi kurdahi}
\IEEEauthorblockA{\textit{Center for Embedded and Cyber-Physical Systems} \\
\textit{University of California, Irvine}\\
Irvine, USA \\
kurdahi@uci.edu}
}
\IEEEaftertitletext{\vspace{-4mm}}
\maketitle
\begin{abstract}
Reliable validation of autonomous driving cyber-physical systems requires testing not only nominal driving behavior but also failures arising from faults in sensing and embedded computation. Physical road testing provides realistic system behavior but is costly and difficult to reproduce, while simulation enables scalable evaluation but cannot fully capture hardware-dependent behavior and fault propagation. Hardware-in-the-loop (HIL) validation bridges this gap by executing the driving stack on real embedded hardware within a simulated environment. However, preparing fault-augmented scenarios for HIL testing remains largely a manual process, limiting test coverage, scalability, and efficiency.

We present FaultGene, an LLM-assisted framework for automated generation of executable fault-augmented tests for HIL fault-resilience validation. Given a baseline Scenic/CARLA scenario and a structured fault specification, FaultGene generates a fault-augmented test case that captures the effects of HIL-level fault injection. During offline refinement, semantic-consistency checks and execution-level feedback are used to correct candidates that fail to preserve the intended scenario context or produce the desired fault-induced degradation, and to construct high-quality supervision for fine-tuning. The resulting generator then produces fault-augmented tests directly at deployment time without requiring manually engineered reference executions or repeated refinement loops.

On held-out scenario–fault combinations, FaultGene improves semantic alignment and executability over one-shot LLM generation across both modular and end-to-end autonomous driving architectures. It also preserves the main sensor- and compute-fault sensitivity trends observed in HIL evaluation, capturing hardware-level fault effects beyond simulation-based verification and falsification alone. Compared with conventional manual fault-scenario engineering, FaultGene reduces estimated fault-scenario preparation time by 72×, enabling scalable HIL-based dependability validation.
\end{abstract}


\section{Introduction}

The rapid advancement of Automated Driving Systems (ADS) has created an urgent need for testing environments that can reliably expose safety-critical failures~\cite{ADtestReview}. Road testing provides high-fidelity evaluation but is costly, time-consuming, difficult to reproduce, and limited in its ability to systematically exercise rare fault conditions. Simulation-based testing offers scalable and reproducible alternatives through controllable scenario generation using platforms such as CARLA~\cite{carla}. Existing probabilistic, constraint-based, and search-based approaches~\cite{scenic,dreossi2019verifaitoolkitdesignanalysis} further expand the scenario space for systematic verification. However, simulation alone cannot fully capture hardware-dependent effects, such as timing variations, scheduling jitter, and embedded computation faults, motivating hardware-in-the-loop (HIL) testing, where the driving stack executes on real hardware within a simulated environment.

Despite these advantages, preparing fault-augmented scenarios for HIL testing remains a practical bottleneck. A fault-oriented test must define both the driving context and the intended sensor or compute fault while preserving scenario semantics and producing meaningful system-level effects. Manually constructing scenarios across fault types, activation times, and severities is labor-intensive and difficult to scale.


Large language models (LLMs) offer a potential solution by translating high-level specifications into executable scenario programs~\cite{zhao2026surveyapplicationlargelanguage}. However, one-shot generation may produce scenarios that violate the intended context, fail to compile or execute, or fail to expose the specified fault meaningfully. These limitations motivate combining LLM-based generation with execution-guided refinement grounded in HIL behavior.

This paper presents \textit{FaultGene}, an LLM-assisted framework for automated generation of fault-augmented HIL test cases. Given a baseline Scenic~\cite{scenic}/CARLA scenario and a structured specification of fault type, activation timing, and severity, FaultGene generates a fault-augmented test while preserving the underlying driving context. The framework integrates structured LLM generation with execution-guided refinement in the Autonomous Driving Digital Twin (ADDT) environment~\cite{anonymous2025addt}. 
HIL executions provide fault-grounded reference behavior and system-level feedback that are used offline to refine generated scenarios and construct supervision for fine-tuning the generator.
After tuning, the generator produces fault-augmented tests in a single pass without reference executions or iterative refinement during deployment.


We evaluate FaultGene on held-out scenario–fault combinations across modular and end-to-end AD architectures. Experimental results show improved scenario validity and executability over one-shot LLM generation, while the generated tests reproduce the main sensor- and compute-fault sensitivity trends observed in ADDT. Compared with conventional manual fault-scenario engineering, FaultGene reduces fault-engineering effort by 72$\times$ per fault type.


In summary, this work makes the following contributions:
\begin{itemize}
\item An LLM-assisted framework for automatically transforming baseline driving scenarios and structured fault specifications into executable fault-augmented test cases;
\item A feedback-driven refinement and tuning mechanism that uses execution feedback as supervision to improve semantic consistency and fault-induced system behavior;
\item An empirical evaluation showing improved scenario validity, ADDT-consistent fault trends, and higher testing efficiency compared with conventional HIL workflows.
\end{itemize}
\section{Related Works}
\label{sec:related}

Digital-twin-based hardware-in-the-loop (HIL) validation enables controlled evaluation of autonomous driving systems under realistic sensing, compute, and communication disturbances. Digital twins model the vehicle, environment, and traffic context~\cite{rasheed2020dt,dt2023}, while HIL co-simulation executes perception, planning, and control on real embedded hardware, exposing timing effects, scheduling jitter, and fault propagation that simulation-only studies may miss~\cite{zhou2023safetyDT,electronics11152462,Abboush2024VirtualTesting}. Fault injection has been used to assess neural-network-based autonomous driving components: APPRAISER~\cite{Taheri2023APPRAISER} and DNN accelerator fault-injection frameworks~\cite{OrdonezControlledFaults} examine model-level vulnerability, while AVFI and Kayotee~\cite{AVFI,Jha2019Kayotee} inject faults across perception, planning, and control in simulated environments. However, these approaches generally do not jointly address fault-oriented scenario generation and realistic embedded execution.

Autonomous-driving scenario generation has explored probabilistic and constraint-based specification, systematic search, and optimization. Scenic~\cite{scenic} enables probabilistic and constraint-based scenario specification, while VerifAI~\cite{dreossi2019verifaitoolkitdesignanalysis} performs simulation-based verification and falsification over parameterized scenario spaces; their combination has been used to identify challenging scenarios~\cite{scenicVerifAI}, alongside optimization- and adversarial-search-based approaches~\cite{adversarialScenario1}. Works have also explored learning scenario representations from driving videos, such as~\cite{scenario2vector}. More recently, LLMs have been applied to driving-scenario description and generation. DriveGPT4~\cite{drivegpt2023} and Talk2Drive~\cite{talk2drive2023} support language-based driving representation and interaction, vision-language approaches derive structured scenarios from driving videos~\cite{scriptgpt2024}, and DriveArena~\cite{lmdrive2024} and LMDrive~\cite{drivearena2024} explore LLM-based scenario expansion and reasoning over traffic interactions and environmental conditions. Nevertheless, these methods typically rely on predefined scenario spaces or text-/simulation-level generation and do not directly construct fault-augmented scenarios from structured hardware-fault specifications or validate their fault-induced effects on real embedded hardware. 
FaultGene addresses this gap by treating fault-augmented test generation as a program-level generation and refinement problem grounded in HIL execution.
\section{HIL-Based Autonomous Driving Validation}
\label{sec:HIL}

\begin{figure*}[t]
    \centering
    \includegraphics[width=0.8\textwidth]{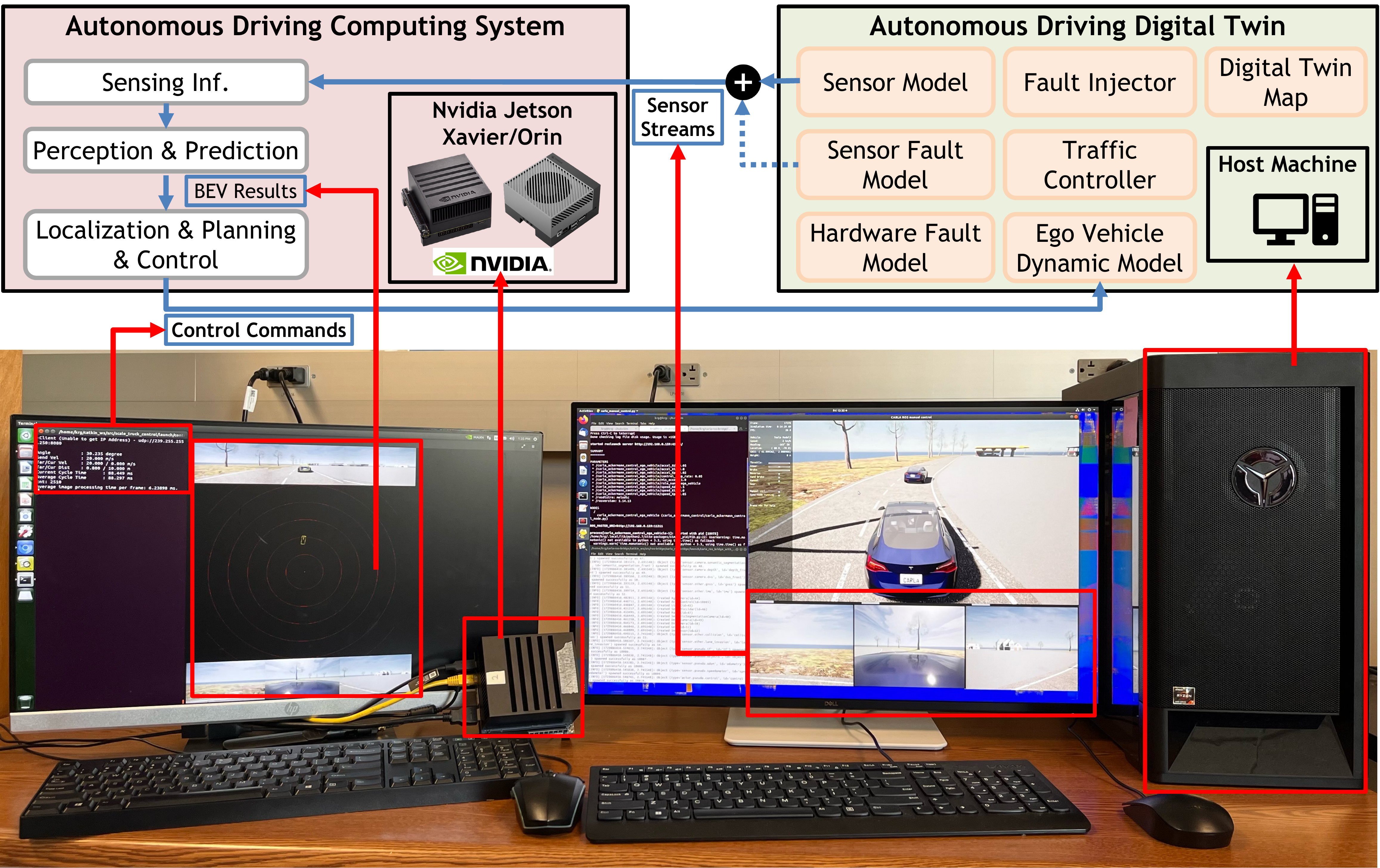}
    \caption{Overview of the ADDT hardware-in-the-loop setup.}
    \label{fig:ADDTframe}
\end{figure*}

In this section, we introduce the Autonomous Driving Digital Twin (ADDT)~\cite{anonymous2025addt}, the HIL validation environment used throughout this work.
\subsection{Hardware-in-the-Loop Setup}

As shown in Fig.~\ref{fig:ADDTframe}, ADDT consists of a simulation system and an AD computing system connected through Ethernet. The simulator streams sensor data to the AD system and receives control commands, enabling closed-loop execution in the virtual environment.

\subsubsection{Digital-Twin Simulation System}

\textbf{Software.}
The simulation system is built on CARLA~\cite{carla}, an open-source AD simulator based on Unreal Engine~\cite{ue}. OpenSCENARIO~\cite{openscenario} and Scenic~\cite{scenic} specify and parameterize the traffic scenarios.

\textbf{Hardware.}
The simulator runs on a workstation with an Intel i7-9800X CPU and an NVIDIA RTX 2080 Ti GPU.

\subsubsection{AD Computing System}
\label{sec:ADCompSys}

%
%

The AD stack (AD architectures are described in Section~\ref{sec:setup}) runs on an NVIDIA Jetson Xavier or Orin, processing six 480p or three 720p camera streams at 10\,Hz and producing control commands at 100\,Hz.

\subsubsection{Testing Scenarios}
\label{sec:ADScenario}

\begin{figure}[t]
    \centering
    \begin{subfigure}{0.88\columnwidth}
        \captionsetup{labelformat=empty, font=scriptsize, aboveskip=-8pt, belowskip=2pt}
        \includegraphics[width=\columnwidth]{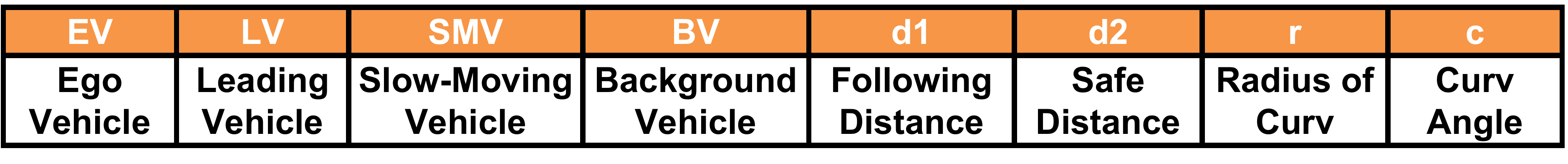}
        \vspace{-9pt}
        \label{fig:Parameters}
    \end{subfigure}

    \begin{subfigure}{0.3\columnwidth}
        \captionsetup{labelformat=empty, font=scriptsize}
        \includegraphics[width=\columnwidth]{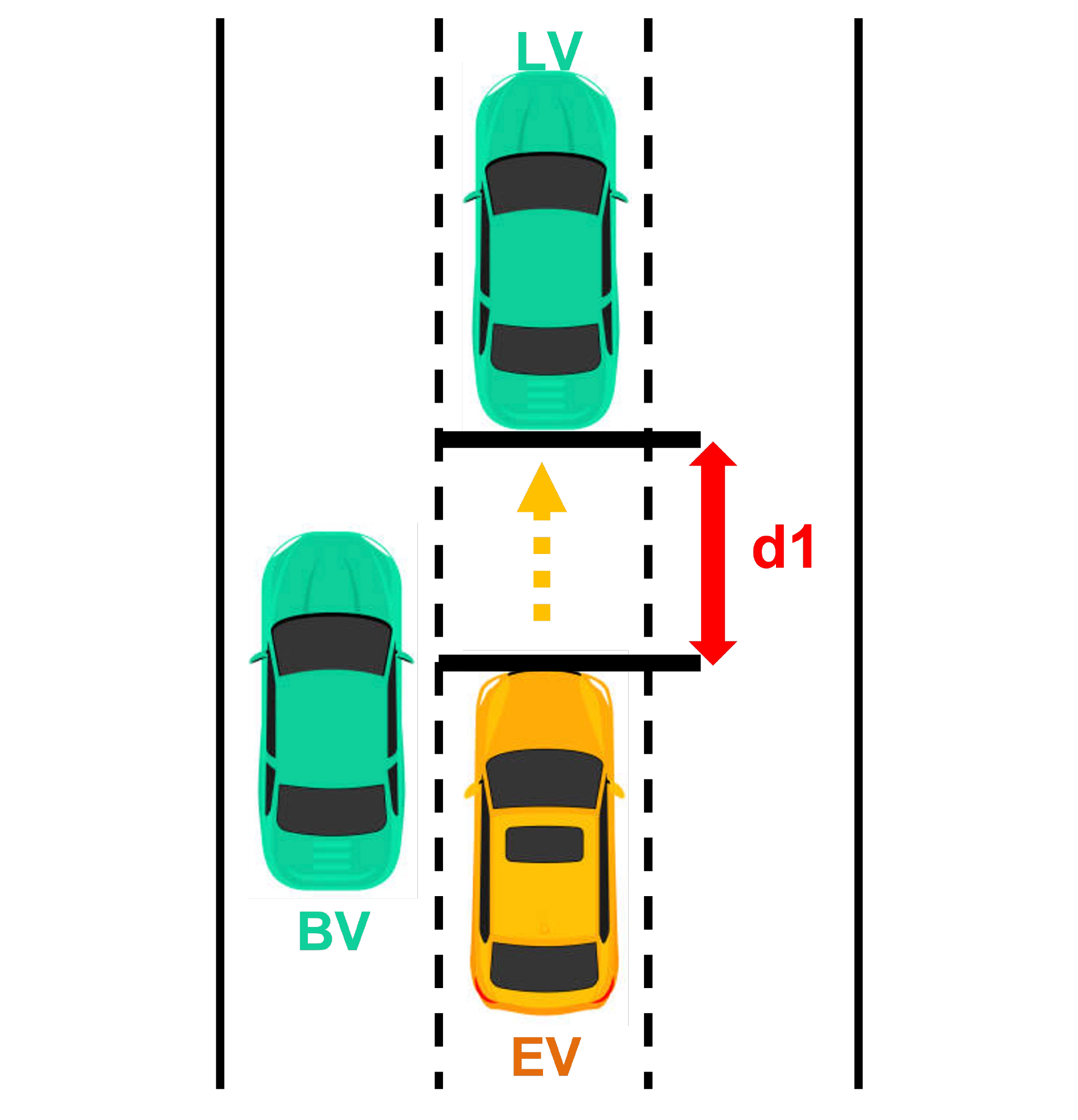}
        \caption{(1) Vehicle following.}
        \label{fig:Scene1}
    \end{subfigure}
    \hfill
    \begin{subfigure}{0.32\columnwidth}
        \captionsetup{labelformat=empty, font=scriptsize}
        \includegraphics[width=\columnwidth]{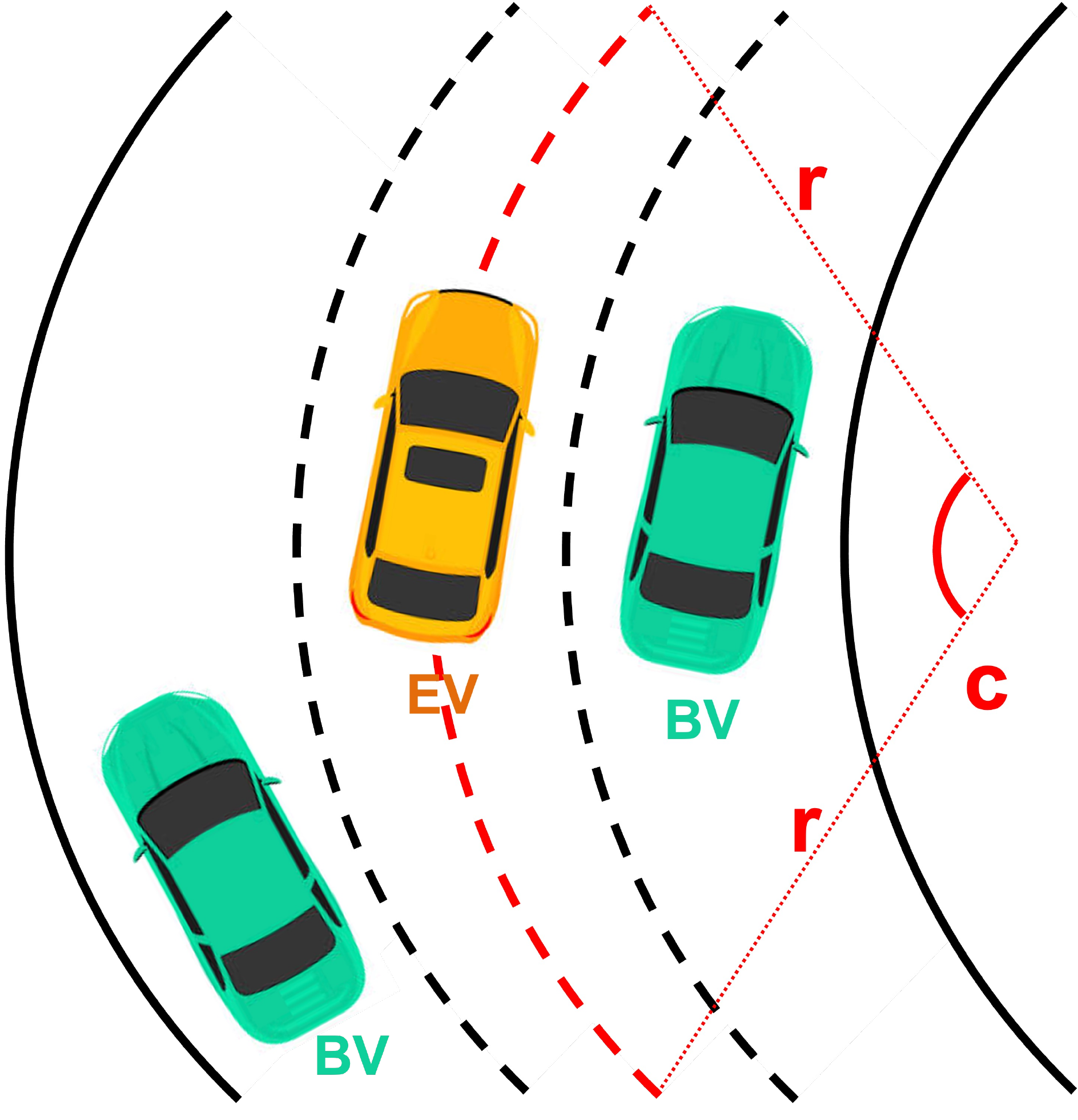}
        \caption{(2) Large-curvature turning.}
        \label{fig:Scene2}
    \end{subfigure}
    \hfill
    \begin{subfigure}{0.3\columnwidth}
        \captionsetup{labelformat=empty, font=scriptsize}
        \includegraphics[width=\columnwidth]{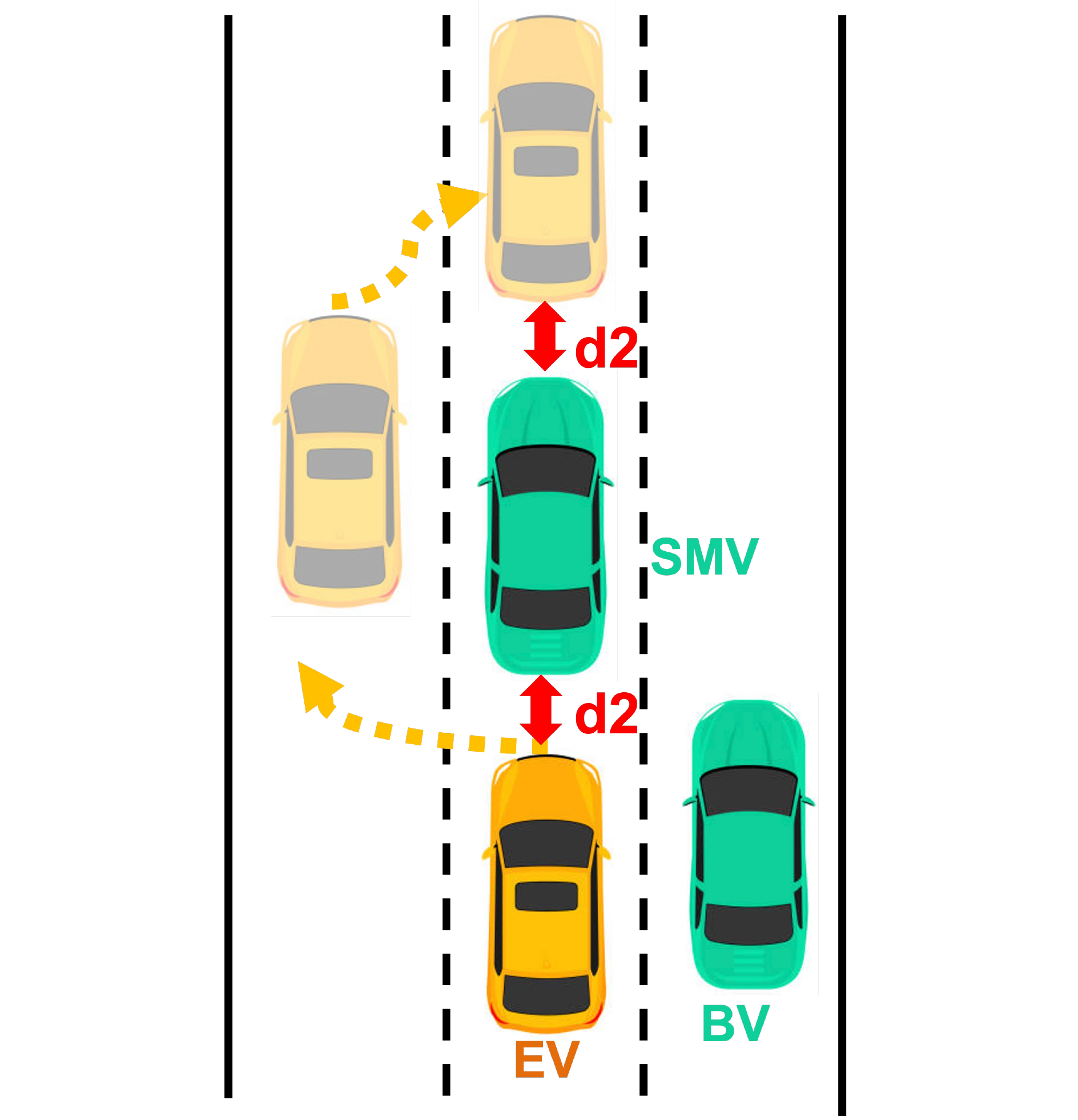}
        \caption{(3) Overtaking.}
        \label{fig:Scene3}
    \end{subfigure}
    \caption{Driving scenarios used for evaluation.}
    \vspace{-10pt}
    \label{fig:Scenarios}
\end{figure}

We evaluate three representative driving scenarios: vehicle following, large-curvature turning, and overtaking. Fig.~\ref{fig:Scenarios} shows their configurations and parameters, including vehicle speeds and inter-vehicle distances.

\subsection{Fault Injection Mechanism}

We evaluate resilience by injecting faults into the sensing and computing paths of the HIL pipeline.

\textbf{Sensor fault injection.}
ADDT models temporal and spatial sensing faults:
\begin{enumerate}
    \item \textbf{Temporal faults:}
    Controlled timestamp delays are introduced into sensor streams. Delays exceeding the perception model's out-of-order tolerance are treated as frame drops.

    \item \textbf{Spatial faults:}
    Position and orientation perturbations are applied to the ground-truth extrinsic parameters to emulate sensor misalignment with controllable severity.
\end{enumerate}

\textbf{Computing hardware fault injection.}
ADDT supports closed-loop CPU and GPU fault injection for studying memory and data-path errors~\cite{wan2024vulnerability,jha2019ml}. CPU faults are injected with LLFI~\cite{lu2015llfi}, while GPU faults are injected with NVBitFI~\cite{tsai2021nvbitfi}. Both tools introduce register-level bit flips at specified locations and values, allowing the resulting errors to propagate through subsequent AD computations. This approach follows prior fault-injection studies~\cite{jha2019ml,wan2024vulnerability}.
\section{FaultGene: Automated Fault-Augmented Test Generation}
\label{sec:faultgene}

\begin{figure*}[t]
\centering
\includegraphics[width=1\textwidth]{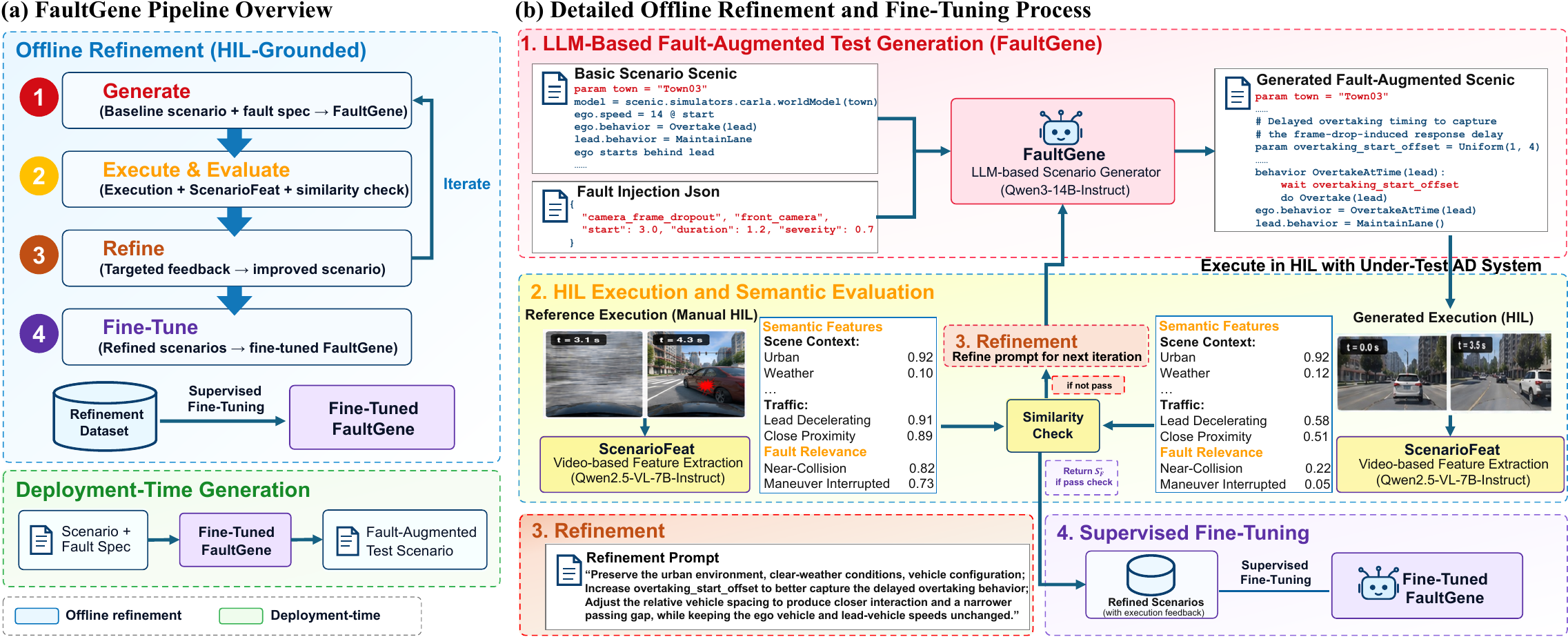}
\caption{
Overview of FaultGene.
(a) Pipeline overview showing offline refinement and deployment-time generation steps.
(b) Detailed offline refinement process, where FaultGene generates fault-augmented Scenic scenarios, compares their executions with manually engineered HIL references using ScenarioFeat, and uses the resulting feedback for refinement and fine-tuning.
}
\label{fig:faultgene_overview}
\end{figure*}


FaultGene is an LLM-based framework for generating executable fault-augmented tests for autonomous-driving fault-resilience validation. An overview of the framework is shown in Fig.~\ref{fig:faultgene_overview}. Given a baseline Scenic/CARLA scenario and a structured fault specification, it produces constrained scenario variants that can be executed in CARLA or through a CARLA-based HIL setup, such as the ADDT HIL environment described in Section~\ref{sec:HIL}.

FaultGene has two phases. During offline framework construction, a prompt-based generator produces candidate fault-augmented Scenic programs, while \textit{ScenarioFeat} extracts semantic attributes from execution videos. Generated executions are compared with manually engineered ADDT HIL references, and detected mismatches guide prompt refinement and supervision construction. After fine-tuning, the deployed generator produces tests in one pass without accessing manually engineered references, execution feedback, or refinement loops.

\subsection{ScenarioFeat: Video-Based Scenario Feature Extraction}
\label{subsec:scenariofeat}

ScenarioFeat is a prompt-engineered module that extracts semantic features from manually engineered reference and FaultGene-generated executions. It uses Qwen2.5-VL-7B-Instruct as its base vision-language model.

For an execution video $V$, ScenarioFeat extracts a predefined semantic feature vector describing the driving context and safety-relevant behavior, including road environment, surrounding-object interactions, obstacle relationships, and vehicle maneuvers such as cut-in behavior:
\begin{equation}
C(V) = [c_1(V), c_2(V), \ldots, c_n(V)],
\end{equation}
where $c_i(V)\in[0,1]$ denotes the predicted confidence score for feature category $i$.

ScenarioFeat is prompt-engineered using representative execution videos and expected feature annotations. It does not generate scenarios; instead, it provides a structured semantic representation for comparing generated executions with manually engineered references.

\subsection{FaultGene: Prompt-Based Scenic Generation}
\label{subsec:faultgene_prompt}

FaultGene is a text-only LLM, instantiated with Qwen3-14B-Instruct, that transforms a baseline Scenic scenario and structured fault specification into a fault-augmented Scenic program. Formally,
\begin{equation}
S_F = \pi_\theta(S_0, F),
\end{equation}
where $S_0$ is the baseline scenario, $F$ is the fault specification, and $\pi_\theta$ is the generator.

The input includes $S_0$, $F$, and constraints requiring valid Scenic syntax and preservation of the original driving context. The fault specification defines the fault type, activation condition, and severity. FaultGene modifies the scenario while preserving its road layout, actor identities, and intended maneuver semantics. Because the output is structured Scenic code rather than free-form text, it can be syntax-checked and executed directly in CARLA or CARLA-based HIL setups such as ADDT.

\subsection{Reference-Guided Similarity and Execution Checks}
\label{subsec:similarity_check}

Each generated scenario $S_F$ is executed through the HIL loop with the under-test AD system and compared with the corresponding manually fault-engineered ADDT HIL execution.
The reference provides offline HIL grounding for supervision construction and is not required at deployment.

Let $V_{\mathrm{ref}}$ and $V_F$ denote the reference and generated execution videos. ScenarioFeat extracts their semantic feature vectors:

\begin{equation}
C(V_{\mathrm{ref}}) =
[\ldots, c_n(V_{\mathrm{ref}})],
\qquad
C(V_F) =
[\ldots, c_n(V_F)].
\end{equation}


The feature mismatch vector is defined as
\begin{equation}
\begin{aligned}
\mathbf{d}_C(V_{\mathrm{ref}},V_F)
&= [c_1(V_{\mathrm{ref}})-c_1(V_F),\ldots,\\
&\quad c_n(V_{\mathrm{ref}})-c_n(V_F)].
\end{aligned}
\end{equation}

A generated execution is semantically inconsistent if
\begin{equation}
|\mathbf{d}_C(V_{\mathrm{ref}},V_F)_i|>\tau_i
\end{equation}
for any feature $i$. The threshold $\tau_i$ controls the tradeoff between semantic preservation and scenario diversity and is used only during offline refinement.

We separately assess fault relevance using near-collision events, inter-vehicle clearance, maneuver interruption, collision outcome, and task completion.
Let $M(S_F)$ denote the resulting metric vector. A scenario is fault-relevant if
\begin{equation}
M(S_F)\in\Omega,
\end{equation}
where $\Omega$ denotes safety-relevant degradation without trivial invalidity, such as deadlock, early termination, or failure to trigger the fault.

Together, these checks determine whether the scenario preserves the intended context, implements the requested fault, and allows it to manifest during execution.

\subsection{Offline Iterative Refinement and Supervised Fine-Tuning}
\label{subsec:iterative_refinement}

Candidates that fail either semantic or fault-relevance checks, or fail to compile or execute successfully, are refined offline.
The mismatch is converted into structured feedback covering Scenic syntax and compilation, physical or semantic validity, fault-specification consistency, and execution-level fault manifestation. Depending on the failure, the feedback may request corrected actor placement, adjusted activation timing, stronger fault severity, or removal of unintended deadlocks and early termination.

Let $\Delta(S_F)$ denote the aggregated semantic and execution mismatch relative to the manually engineered reference. Starting from $S_F^{(0)}$, the generator produces the next candidate as
\begin{equation}
S_F^{(k+1)}
=
\mathrm{Refine}\!\left(
S_0,F,S_F^{(k)},\Delta(S_F^{(k)})
\right),
\end{equation}
until
\begin{equation}
|\mathbf{d}_C(V_{\mathrm{ref}},V_F^{(k)})_i|
\le \tau_i,\quad \forall i,
\end{equation}
and
\begin{equation}
M(S_F^{(k)})\in\Omega.
\end{equation}

Here, $\mathrm{Refine}(\cdot)$ denotes rule-guided LLM regeneration using the original baseline scenario, fault specification, current candidate, and structured feedback prompts. It is not a reinforcement-learning policy or an external optimizer that directly edits the Scenic program.

The resulting scenario $S_F^\star$ is retained as the corrected target for $(S_0,F)$. Repeating this process over $N$ framework-construction examples yields
\begin{equation}
\mathcal{D}_{\mathrm{ref}}
=
\left\{
\left(S_0^{(j)},F^{(j)},S_F^{\star(j)}\right)
\right\}_{j=1}^{N}.
\end{equation}

We then fine-tune the original generator $\pi_\theta$ on $\mathcal{D}_{\mathrm{ref}}$ using supervised fine-tuning (SFT). Each training example uses the baseline scenario and fault specification $(S_0,F)$ as input and the corresponding refined Scenic program $S_F^\star$ as the target. The trainable LoRA parameters are optimized with teacher-forced causal language modeling:
\begin{equation}
\mathcal{L}_{\mathrm{SFT}}
=
-\sum_{t=1}^{|S_F^\star|}
\log P_\theta\!\left(
s_t^\star
\mid
S_0,F,s_{<t}^\star
\right),
\end{equation}
where $s_t^\star$ is the $t$-th token of $S_F^\star$ and $s_{<t}^\star$ denotes its preceding tokens. The baseline scenario and fault specification provide the conditioning context, while the loss is computed over the target Scenic program.

Although the loss is evaluated token by token, the model is conditioned on the complete $(S_0,F)$ pair and the preceding program context. Across diverse training examples, the shared parameters learn reusable transformations, including which scenario elements to preserve, how to encode the specified fault, and how to adjust actor timing and interactions to produce valid and fault-relevant Scenic programs. The objective therefore learns the conditional mapping
\begin{equation}
(S_0,F)\longmapsto S_F^\star.
\end{equation}

The refined target $S_F^\star$ is used rather than the manually engineered HIL reference because it converts execution-level behavioral supervision into an input-aligned program target. Since $S_F^\star$ is generated from the $(S_0,F)$ pair and refined against the reference, SFT can directly learn the desired program transformation, following the broader paradigm of using validated or refined generated outputs as supervision~\cite{wang2023selfinstruct,huang2023large}.

This SFT process absorbs the offline corrections into the trainable LoRA parameters, producing the finalized generator $\pi_{\theta^\star}$. Its generalization is evaluated on held-out scenario--fault combinations excluded from refinement and fine-tuning.

\subsection{Deployment-Time Fault-Augmented Test Generation}
\label{subsec:deployment_time}

After fine-tuning, the deployed model generates
\begin{equation}
S_F^{\mathrm{deploy}}
=
\pi_{\theta^\star}(S_0,F),
\end{equation}
where $\theta^\star$ denotes the fine-tuned model parameters and $(S_0,F)$ is a new baseline scenario--fault specification pair.
The generated Scenic program is syntax-checked and compiled for execution with CARLA or a CARLA-based HIL setup.

Because the offline corrections are absorbed through fine-tuning, deployment does not require a manually engineered reference, ScenarioFeat comparison, execution feedback, or repeated refinement. FaultGene therefore generates fault-augmented tests in one pass.
\section{Experimental Setup}
\label{sec:setup}

\begin{table*}[t]
\centering
\caption{Semantic alignment, compilation success, and execution success for modular and end-to-end systems.}
\label{tab:modular_vs_e2e}
\small
\begin{tabular}{llccc}
\toprule
\textbf{System} & \textbf{Method} & \textbf{Semantic Alignment} & \textbf{Compilation} & \textbf{Execution Success} \\
 &  & \textbf{Score*} & \textbf{Success(\%)} & \textbf{Among Compiled(\%)} \\
\midrule
Modular  & One-shot LLM & 0.45 & 68 & 81 \\
Modular  & Fine-tuned FaultGene Generator & 0.89 & 96 & 93 \\
\midrule
End-to-end & One-shot LLM & 0.43 & 71 & 74 \\
End-to-end & Fine-tuned FaultGene Generator & 0.86 & 94 & 90 \\
\bottomrule
\end{tabular}
\begin{flushleft}
\footnotesize
\textit{*The semantic alignment score is computed as the average feature agreement between the generated and reference executions.}
\end{flushleft}
\end{table*}

\textbf{Simulation Environment.}
We use CARLA~\cite{carla}, an open-source platform for the development and validation of autonomous driving systems. CARLA provides a realistic physics engine and high-fidelity environmental rendering, enabling the generation of reproducible driving scenarios for our evaluation.

\textbf{Scenario Description.}
We use Scenic~\cite{scenic} to describe driving scenarios. Its Python-like syntax provides a structured representation that is well suited for LLM-based scenario generation and prompt engineering. Scenic also integrates directly with CARLA and supports the specification of complex vehicle configurations, traffic interactions, and environmental conditions, making it suitable for program-level scenario generation.

\textbf{Autonomous Driving Systems Under Test.}
We evaluate FaultGene on two complementary autonomous driving architectures: a modular stack that explicitly separates perception, planning, and control, and an end-to-end system that directly maps sensor observations to driving commands. This distinction allows us to evaluate FaultGene across different system organizations and fault-propagation paths.

\textit{Modular AD stack:}
The modular system consists of a FastBEV~\cite{FastBEV} perception module, an MLP-based planner trained on nuPlan~\cite{caesar2021nuplan}, and a PID controller. The modules communicate through ROS~\cite{ROS} and execute on the embedded HIL platform, enabling evaluation of fault propagation across modules.

\textit{End-to-end PilotNet:}
The end-to-end system uses PilotNet~\cite{pilotnet}, a CNN-based model  designed for lane-keeping that directly maps front-camera images to steering commands. We use PilotNet as a simple representative of end-to-end driving, providing a complementary architecture for evaluating FaultGene through an end-to-end driving pipeline without explicit perception, planning, and control modules.

\textbf{Dataset split and evaluation protocol.}
We construct the refinement dataset from 600 baseline scenario--fault combinations and reserve 200 disjoint combinations for held-out evaluation. No held-out is used during refinement or fine-tuning.

\textbf{Fine-Tuning Setup.}
FaultGene is fine-tuned using QLoRA-based supervised fine-tuning~\cite{dettmers2023qlora}. The training set consists of 600 baseline scenario--fault combinations, each paired with its corresponding refined Scenic program $S_F^\star$ from refinement. 
During fine-tuning, the pretrained Qwen3-14B base model is kept frozen and loaded in 4-bit quantized form, while trainable LoRA adapters are optimized on these input--target pairs.
\section{Results and Evaluation}
In this section, we evaluate FaultGene on held-out scenario–fault combinations from three perspectives: scenario validity and executability, consistency with known fault-sensitivity trends, and overall testing efficiency.

\begin{table}[t]
\centering
\caption{Comparison of sensor-fault sensitivity trends between ADDT and
FaultGene under different camera frame drop rates. Values report mission
success rate (\%).}
\label{tab:sensor_consistency}
\small
\resizebox{\columnwidth}{!}{%
\begin{tabular}{lllcccc}
\toprule
\textbf{System} & \textbf{Source} & \textbf{Scenario} &
\textbf{1\%} & \textbf{2\%} & \textbf{5\%} & \textbf{10\%} \\
\midrule
\multirow{6}{*}{Modular}
& \multirow{3}{*}{ADDT}
& Large-curv. turning & 100.00\% & 99.58\% & 96.25\% & 88.75\% \\
&& Vehicle following & 99.58\% & 99.58\% & 97.92\% & 94.16\% \\
&& Overtaking & 97.08\% & 87.50\% & 73.75\% & 54.58\% \\
\cmidrule{2-7}
& \multirow{3}{*}{FaultGene}
& Large-curv. turning & 99.58\% & 98.60\% & 94.83\% & 86.18\% \\
&& Vehicle following & 97.92\% & 96.05\% & 95.91\% & 92.37\% \\
&& Overtaking & 94.83\% & 85.25\% & 71.69\% & 62.33\% \\
\midrule
\multirow{4}{*}{End-to-end}
& \multirow{2}{*}{ADDT}
& Lane following & 100.00\% & 100.00\% & 94.16\% & 90.33\% \\
&& Large-curv. turning & 98.39\% & 98.39\% & 92.58\% & 84.60\% \\
\cmidrule{2-7}
& \multirow{2}{*}{FaultGene}
& Lane following & 100.00\% & 99.37\% & 95.27\% & 88.45\% \\
&& Large-curv. turning & 99.17\% & 98.12\% & 91.86\% & 82.56\% \\
\bottomrule
\end{tabular}
}
\end{table}

\subsection{Scenario Validity, Executability, and the Effect of Iterative Refinement and Fine-Tuning}
\label{subsec:validity_refinement}

We first evaluate whether fine-tuned FaultGene generator produces valid and executable tests and how iterative refinement and fine-tuning affect semantic alignment and execution quality. A test is considered valid if it compiles successfully in Scenic and executes without runtime failure. Table~\ref{tab:modular_vs_e2e} summarizes the results for a modular autonomous driving stack and an end-to-end system based on PilotNet~\cite{pilotnet}, demonstrating that the same framework supports different AD architectures.

\textbf{Compilation validity and executability.}
For the modular stack, one-shot generation without refinement achieves only 68\% compilation success, with failures mainly caused by syntax errors or invalid parameter combinations. Among compiled scenarios, 81\% complete closed-loop execution, while the remaining failures primarily result from physically infeasible initializations, including overlapping vehicles and deadlocked traffic configurations. With fine-tuned FaultGene, compilation success rises to 96\% and execution success to 93\%, indicating that execution-guided correction and fine-tuning resolve many invalid cases.

A similar pattern appears for the end-to-end system. Without refinement, 71\% of generated scenarios compile successfully, and 74\% of compiled scenarios complete execution. With fine-tuned FaultGene, compilation and execution success increase to 94\% and 90\%, respectively, showing that the refinement benefits both modular and end-to-end systems.

\begin{table*}[t]
\centering
\caption{Estimated deployment-stage efficiency for 1,000 total scenario runs and 20 fault types on the same HIL platform.}
\label{tab:efficiency}
\small
\begin{tabular}{lccccc}
\toprule
\textbf{Metric} & \textbf{Manual ADDT} & \textbf{FaultGene + ADDT*}
& \textbf{FaultGene benefit} \\
\midrule
Fault-scenario preparation
& Manual engineering
& Single-pass generation
& Automated preparation \\
Preparation per fault type
& 12 h
& \textbf{10 min}
& \textbf{72$\times$ preparation-time speedup} \\
Preparation for 20 fault types
& 240 h
& \textbf{3.33 h}
& \textbf{72$\times$ preparation-time speedup} \\
Setup + execution for 1,000 runs
& 200 h
& 200 h
& Unchanged HIL run time \\
Total campaign time
& 440 h
& \textbf{203.33 h}
& \textbf{2.16$\times$ speedup; 53.8\% less time} \\
\bottomrule
\end{tabular}
\begin{flushleft}
\footnotesize
\textit{Assumptions:} Each scenario run requires 2 min of setup and 10 min of HIL execution. Manual fault-scenario preparation is estimated at 12 h per fault type, while FaultGene requires 10 min per fault type for single-pass generation using the fine-tuned generator. Thus, for $R$ total runs and $K$ fault types, $T_{\mathrm{manual}}=0.2R+12K$ and $T_{\mathrm{FG}}=0.2R+K/6$ hours.

\textit{*FaultGene generation is followed by execution on the same ADDT HIL platform.}
\end{flushleft}
\end{table*}

\begin{figure*}[t]
    \centering
    \includegraphics[width=0.8\textwidth]{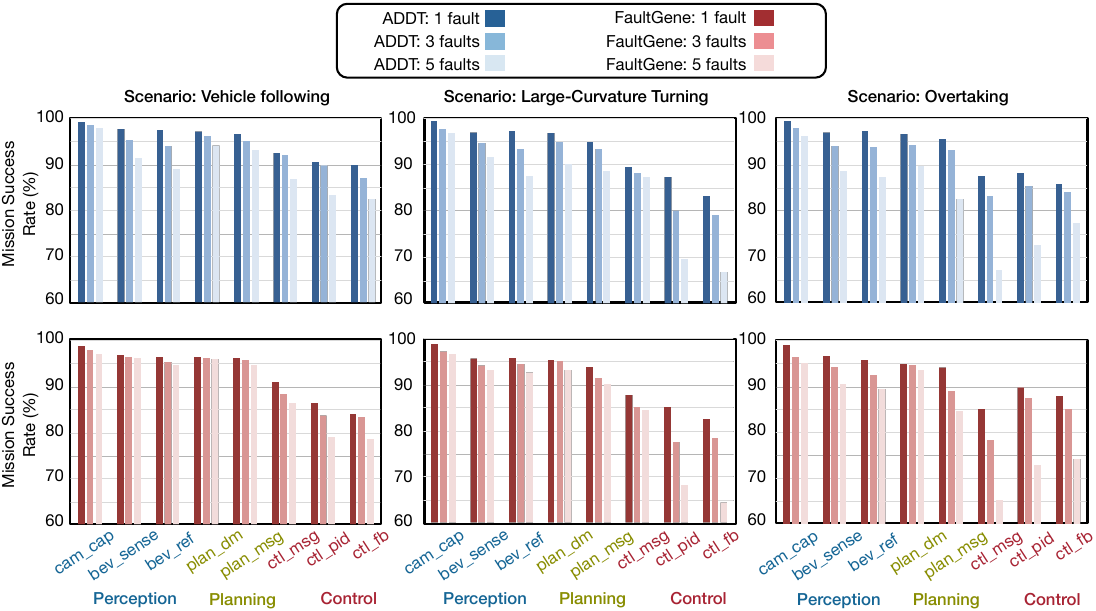}
    \caption{Comparison of compute-fault sensitivity trends between ADDT and FaultGene on the modular AD system across 3 driving scenarios. For each scenario, mission success rate is shown under bit-flip faults injected into perception, planning, and control nodes with 1, 3, and 5 injected faults.}
    \label{fig:bit_flip_compare}
\end{figure*}

\textbf{Semantic alignment.}
ScenarioFeat measures whether generated fault-augmented scenarios preserve the intended behavior by scoring predefined scenario feature categories. As shown in Table~\ref{tab:modular_vs_e2e}, one-shot generation achieves semantic alignment scores of 0.45 and 0.43 for the modular and end-to-end systems, respectively. Fine-tuned FaultGene improves these scores to 0.89 and 0.86, indicating that refinement and fine-tuning improve both executability and preservation of the intended scenario structure and behavior.


\subsection{Consistency with ADDT Fault Trends}
\label{subsec:addt_consistency}

A key question is whether FaultGene-generated tests preserve the system-level fault-sensitivity trends previously observed in ADDT. We therefore compare the qualitative sensor- and compute-fault behaviors exposed by FaultGene with those reported in ADDT. The objective is not to reproduce identical mission success values, but to preserve relative vulnerability patterns across scenarios and software modules.

\textbf{Sensor-fault consistency.}
Table~\ref{tab:sensor_consistency} compares ADDT and FaultGene under different camera frame-drop rates. In both evaluations, the impact of sensor faults is scenario dependent: vehicle following is comparatively robust, large-curve turning shows moderate degradation, and overtaking is the most sensitive. Although the absolute success rates differ, FaultGene preserves this overall ordering as the drop rate increases.

In particular, overtaking degrades more rapidly than vehicle following and large-curve turning, reflecting the stronger dependence of interaction-heavy maneuvers on temporally consistent sensing. This agreement indicates that FaultGene preserves the scenario-dependent sensor-fault behavior observed in ADDT.

The same consistency is observed for the end-to-end system: FaultGene follows the ADDT degradation trends for both normal lane keeping and large-curve turning across increasing drop rates. Since PilotNet is designed primarily for lane-keeping, the end-to-end evaluation focuses on these two scenarios.

\textbf{Compute-fault consistency.}
We evaluate compute-fault consistency on the modular architecture.
Fig.~\ref{fig:bit_flip_compare} compares compute-fault sensitivity across the three  scenarios and multiple fault locations. FaultGene preserves the main trend observed in ADDT: faults injected into later pipeline stages cause greater mission-success degradation, while perception faults are comparatively less damaging. This pattern is especially clear in large-curvature turning and overtaking, where control-node faults cause the sharpest degradation as the number of injected faults increases.

Across modules, perception faults generally produce smaller mission-success drops, planning faults have an intermediate effect, and control faults are the most harmful because they directly affect vehicle actuation. Although the absolute values differ, FaultGene follows the same progression across scenarios, modules, and fault counts, indicating that the generated tests preserve the fault-propagation patterns observed under ADDT HIL execution.

\textbf{Scenario-specific sensitivity.}
FaultGene also preserves the scenario-level conclusions reported in ADDT. Overtaking shows greater sensitivity to planning faults because lane-change-like maneuvers depend strongly on timely decision making. Large-curvature turning is particularly vulnerable to control faults, where late-stage actuation errors quickly cause failure. Vehicle following is more stable overall but retains the same module-level ranking.



\subsection{Testing Efficiency}
\label{subsec:efficiency}

Table~\ref{tab:efficiency} estimates the deployment-stage efficiency of FaultGene for a campaign of 1,000 scenario runs across 20 fault types. FaultGene reduces fault-scenario preparation from 12~h to 10~min per fault type, yielding a 72$\times$ preparation-time speedup. With the same HIL setup and execution time for both workflows, the estimated total campaign time decreases from 440~h to 203.33~h, corresponding to a 2.16$\times$ speedup and 53.8\% reduction in total time.

ADDT and FaultGene use the same hardware configuration, including a simulation workstation, an NVIDIA Jetson Xavier/Orin, and standard HIL networking, with an estimated hardware cost below \$5{,}000. In contrast, an autonomous driving test vehicle, such as a Tesla Model Y, typically costs more than \$50{,}000, excluding insurance, maintenance, calibration, and safety personnel. Thus, FaultGene retains ADDT's approximately 10$\times$ hardware-cost advantage over physical testing while reducing deployment-stage test-preparation effort. Compared with simulation-only testing, it executes the AD stack on real embedded hardware, enabling evaluation of hardware-dependent effects such as compute faults, timing behavior, and fault propagation.

\section{Conclusion}

This paper presented FaultGene, an LLM-based framework for generating fault-augmented tests for HIL resilience validation. Given a baseline Scenic/CARLA scenario and structured fault specification, FaultGene generates executable tests for closed-loop evaluation across modular and end-to-end AD architectures. Manual ADDT references and execution feedback are used only for offline supervision construction and fine-tuning, while deployment requires neither reference executions nor iterative refinement. On held-out scenario--fault combinations, FaultGene improves semantic alignment, compilation, and execution success over one-shot generation while preserving the main HIL sensor- and compute-fault sensitivity trends.

FaultGene reduces manual Scenic editing and fault-augmented tests construction to providing a baseline scenario and fault specification. Although evaluated on ADS, the framework can be adapted to other CPS by replacing the scenario representation, execution metrics, and fault-injection interfaces with domain-specific counterparts. Similarly, although the current evaluation uses LLFI and NVBitFI, FaultGene is not tied to these tools and can incorporate other fault-injection backends. A current limitation is that timing effects are evaluated after generation rather than incorporated into the generation objective. Future work will integrate HIL timing measurements and broader fault-injection backends into the feedback process to expose both safety- and timing-critical failures.
\newpage
\bibliographystyle{IEEEtran}
\bibliography{sn-bibliography}
\end{document}